\documentclass[%
aps,%
preprint,%
showpacs,
nofootinbib,
superscriptaddress
]{revtex4}
\usepackage{epsfig}
\usepackage{amsmath}
\newcommand{\lesssim}{ \mathop{}_{\textstyle \sim}^{\textstyle <} }

\def\be{\begin{equation}}
\def\ee{\end{equation}}

\begin{document}

\preprint{CERN-PH-TH/2006-069}

\title{The $\nu$MSM, Inflation, and Dark Matter}

\author{Mikhail Shaposhnikov}
\affiliation{
Institut de Th\'eorie des Ph\'enom\`enes Physiques,
Ecole Polytechnique F\'ed\'erale de Lausanne,
CH-1015 Lausanne, Switzerland}
\affiliation{Theory Division, CERN, CH-1211 Geneve 23, Switzerland}

\author{Igor Tkachev}
\affiliation{Theory Division, CERN, CH-1211 Geneve 23, Switzerland}

%\date{\today}
\date{April 27, 2006}

\begin{abstract}

We show how to enlarge the $\nu$MSM  (the minimal extension of the
standard model by three right-handed neutrinos) to incorporate
inflation and provide a common source for electroweak symmetry
breaking and for right-handed neutrino masses. In addition to
inflation, the resulting theory  can explain simultaneously dark
matter and the baryon asymmetry of the Universe; it is consistent
with experiments on neutrino oscillations and with all astrophysical
and cosmological constraints on sterile neutrino as a dark matter
candidate. The mass of inflaton can be much smaller than the
electroweak scale.

\end{abstract}


\maketitle

%%%%%%%%%%%%%%%%%%%%%%%%%%%%%%%%%%%%%%%%%%%%%%%%%%%%%%%%%%%%%%%%%%%%%%%%
% Introduction
%%%%%%%%%%%%%%%%%%%%%%%%%%%%%%%%%%%%%%%%%%%%%%%%%%%%%%%%%%%%%%%%%%%%%%%%
%{\em Introduction.}--- 
%

In refs. \cite{Asaka:2005an,Asaka:2005pn} it has been shown that a 
simple renormalizable extension of the Minimal Standard Model (MSM),
containing three right-handed neutrinos $N_I$ of masses smaller than
the electroweak scale, called $\nu$MSM, can explain simultaneously
dark matter and the baryon asymmetry of the Universe, being
consistent with neutrino masses and mixings observed
experimentally\footnote {We do not include here the LSND
anomaly~\cite{Aguilar:2001ty}, which will be tested in the near
future~\cite{MiniBooNE}.}.  The Lagrangian of the $\nu$MSM is 
\begin{eqnarray}
  {\cal L}_{\nu \rm{MSM}}
  = {\cal L}_{\rm{MSM}} + \bar N_I i \partial_\mu \gamma^\mu N_I
  - F_{\alpha I} \,  \bar L_\alpha N_I \Phi
  - \frac{M_I}{2} \; \bar {N_I^c} N_I + \rm{h.c.} \,,
  \label{lagr}
\end{eqnarray}
where ${\cal L}_{\rm{MSM}}$ is the Lagrangian of the MSM, 
$\Phi$ and $L_\alpha$ ($\alpha=e,\mu,\tau$) are respectively the
Higgs and lepton doublets, $F$ is a matrix of Yukawa coupling
constants, and $M_I$ are the Majorana masses. The Majorana mass
matrix is taken to be real and diagonal.

The $\nu$MSM has two essential drawbacks. On the cosmological side,
it does not explain the main features of the Universe, such as its
homogeneity and isotropy on large scales and the existence of
structures on smaller scales; these are believed to be coming from
inflation
\cite{Starobinsky:1979ty}--\cite{Albrecht:1982wi}\footnote{
Note that the present day 
accelerated expansion of the Universe can be incorporated in MSM or
$\nu$MSM by adding a cosmological constant.}. On the theoretical
side, it remains unclear why the potentially different energy scales,
related to the electroweak symmetry breaking and to the Majorana
neutrino masses, could be of the same order of magnitude. The aim of
the present note is to show that a simple extension of the $\nu$MSM
by a real scalar field (inflaton) with scale-invariant couplings (on
the classical level) provides a  possibility to have inflation and to
fix the scales of the $\nu$MSM\footnote{Though our motivation is
similar to that of \cite{Davoudiasl:2004be}, the model we propose and
its physics are entirely different.}. Moreover, we will show that the
coupling of the inflaton to sterile neutrino gives rise to a novel
mechanism of dark matter production. In order not to let the number
of different names proliferate, we will be using in what follows the
same name ($\nu$MSM) for the theory with the inflaton.

We propose the following Lagrangian to describe the physics at 
energies below the Planck scale:
\begin{eqnarray}
  {\cal L}_{\nu \rm{MSM}}\rightarrow
   {\cal L}_{\nu \rm{MSM}[M\rightarrow 0]} + 
  \frac{1}{2}(\partial_\mu\chi)^2 -
  \frac{f_I}{2} \; \bar {N_I}^c  N_I \chi + \rm{h.c.} - V(\Phi,\chi)
   \,,
  \label{lagr1}
\end{eqnarray}
where the first term is the $\nu$MSM Lagrangian with all dimensionful
parameters (Higgs and Majorana masses)  put to zero, $\chi$ is a real
scalar field (inflaton), and  $f_I$ are  Majorana-type Yukawa
couplings. We parametrize the most general scale-invariant potential 
$V_s(\Phi,\chi)$ as follows
\be
V_s(\Phi,\chi) =
\lambda\left(\Phi^\dagger\Phi-\frac{\alpha}{\lambda}\chi^2\right)^2+
\frac{\beta}{4}\chi^4~,
\label{pot}
\ee
where $\lambda,~\alpha$ and $\beta$ are the scalar coupling constants,
which we take to be positive. We will assume that the scale
invariance is explicitly broken on the classical level in the
inflaton sector only, so that
\be
V(\Phi,\chi)= -\frac{1}{2} m_\chi^2 \chi^2 + V_s(\Phi,\chi)~.
\label{pottot}
\ee
This potential has a symmetry $\chi \to -\chi$ which could lead to a
cosmological domain wall problem \cite{Zeldovich:1974uw}. However,
there are many different ways to solve this problem in inflationary
cosmology, the simplest one is just to add to (\ref{pottot}) a cubic
term $\sim \chi^3$.

The requirement of scale invariance looks (and is) rather ad-hoc
since this symmetry is broken by quantum corrections. We cannot
provide any further motivation for this choice besides the one given
already in ref. \cite{Buchmuller:1990pz} (see also references
therein). From a phenomenological side, it is this requirement that
ensures the same source for the scale of electroweak symmetry
breaking and for Majorana masses of sterile neutrinos. An explicit
breaking of this symmetry by the inflaton mass term is also
essential: if $m_\chi^2=0$ the electroweak symmetry breaking by
radiative corrections only is impossible, because of the large mass
of $t$-quark  \cite{Buchmuller:1990pz}. From (\ref{pot},\ref{pottot})
one gets the relations between the vev of the inflaton field, its
mass $m_I$ and the Higgs mass $m_H$: $\langle \chi \rangle \simeq
m_H/2\sqrt{\alpha}$,  $m_I \simeq m_H \sqrt{\beta/2\alpha}$. In what
follows we will choose $\alpha > \beta/2$. In this case the inflaton
is lighter than the Higgs boson, $m_I<m_H$.

Let us discuss different constraints on the parameters of this model
coming from successful (chaotic) inflation scenario
\cite{Linde:1983gd}. The inflaton potential must be sufficiently flat
so as not to produce too large density fluctuations. In our case the
flat direction is given by
$\Phi^\dagger\Phi=\frac{\alpha}{\lambda}\chi^2$, since the constant
$\lambda \sim 1$ must be large enough to have a Higgs boson of mass
${\cal O}(100)$ GeV. Along this direction the potential is simply $V
\propto \frac{\beta}{4}\chi^4$. The constant $\beta$ can now be fixed
from the requirement to give correctly the amplitude of adiabatic
scalar perturbations\footnote{The pure $\chi^4$ inflaton potential
with minimal coupling to gravity is disfavoured by the WMAP3 data
\cite{Spergel:2006hy}, producing too large tensor fluctuations.
However, allowing for  non-minimal couplings (see also the comment at
the end of the paper) can bring this potential in agreement with the
data \cite{Hwang:1998mx,Komatsu:1999mt}.} observed by WMAP
\cite{Spergel:2006hy}. With the use of general expressions given, for
instance, in \cite{Lyth:1998xn}, one gets $\beta \simeq 2.6\times
10^{-13}$. The flatness of the potential must not be spoiled by
radiative corrections from the loops of the particles of the Standard
Model and sterile neutrinos. This requirement gives 
$  
\alpha \lesssim 3 \times 10^{-7},~~f_I 
\lesssim 2 \times 10^{-3}~.  
$

Another constraint could come from the requirement to have successful
baryogenesis. After the end of inflation the Universe is reheated up
to a certain temperature $T_r$, which must be considerably larger
than the freezing temperature of anomalous electroweak fermion number
non-conservation $T \simeq 130$ - $190$ GeV \cite{Burnier:2005hp} to
allow sphaleron processes to convert the lepton asymmetry created in
sterile--active neutrino transitions to baryon asymmetry
\cite{Kuzmin:1985mm,Akhmedov:1998qx,Asaka:2005pn}.  In our model, the
transfer of inflaton energy to the fields of the Standard Model goes
through the inflaton--Higgs coupling, proportional to the parameter
$\alpha$. The energy in the Higgs field is then quickly distributed
among all other fields of the MSM, since the typical coupling
constants are quite large. The evolution of the energy $\rho =
\frac{\pi^2 g^*}{30}T^4$ of the MSM particles can be found from
equation
\be
\frac{\partial \rho}{\partial t} + 4 H \rho = R~,
\ee
where $H$ is the Hubble constant and  $R$ is the energy transfer rate
from inflaton oscillations to the Higgs field, which can be 
found using the approach of Ref. \cite{musher,Micha:2004bv}
\be
R \sim \frac{1}{\lambda} \alpha^2 \omega \chi^4~.
\ee
Here  $\omega^2 \sim \beta\chi^2$ is the typical
frequency of inflaton oscillations. Taking into account that the
Universe  expands as 
radiation dominated after inflation (since we assumed
that $m_\chi \ll \sqrt{\beta}M_{\rm{Pl}}$) one gets 
\be
T_r \sim M_{\rm{Pl}}
\left(\frac{\alpha^2}{g^*\lambda}\right)^{1/4}~.
\ee
For $\alpha > \beta \simeq 10^{-13}$ this temperature exceeds 
greatly the electroweak scale, as required.

Now, we are coming to the question of dark matter abundance in this
model. The lightest sterile neutrino $N_1$, being sufficiently
stable, plays the role of dark matter in $\nu$MSM. It can be created 
in active--sterile neutrino oscillations \cite{Dodelson:1993je} or
through the coupling to the inflaton
\cite{Boyarsky:2005us,Asaka:2006ek}. We will assume that the Yukawa
constants $F_{\alpha 1}$ are too small to make the first mechanism
operative, $F_{\alpha 1} \lesssim 10^{-12}$ 
\cite{Asaka:2005an,Asaka:2005pn} and estimate the second effect
only. 

Owing to the Higgs--inflaton mixing, the inflaton with a mass $300$
MeV $\lesssim m_I \lesssim m_H$ is in thermal equilibrium down to
rather small temperatures $T \ll m_I$, thanks to reactions $\chi
\leftrightarrow e^\dagger e,~ \chi \leftrightarrow \mu^\dagger \mu$,
etc\footnote{If inflaton mass is larger than $500$ GeV one can show
that the inflaton does not equilibrate. In this case the computation of
sterile dark matter abundance requires a detailed study of fermionic
preheating, similar to \cite{Giudice:1999fb,Greene:2000ew}.}. 
This range of masses
corresponds to $1\times 10^{-13} \lesssim \alpha \lesssim 5 \times
10^{-8}$ and to the inflaton vev in the interval  $4\times 10^5$
GeV--$3\times 10^8$ GeV (we took $m_H = 200$ GeV for
numerical estimates). Sterile neutrinos are produced in inflaton
decays mainly at $T\simeq m_I$, and their distribution function
$n(p,t)$ ($p$ is the momentum of the sterile neutrino and $t$ is
time) can be found from the solution of the kinetic equation
\be
\frac{\partial n}{\partial t} - H p \frac{\partial n}{\partial p}=
\frac{2 m_I \Gamma}{p^2}\int_{p+m_I^2/4p}^\infty n_I(E)dE~,
\label{kin}
\ee
where the inverse decays $\chi \leftarrow N_1N_1$ are neglected, $H$ is
the Hubble constant, $E$ is the inflaton energy, $n_I(E)$ is the
inflaton distribution, $\Gamma = f_1^2 m_I/16\pi$ is the partial
width of the inflaton for $\chi \rightarrow N_1N_1$ decay. For the
case when the effective number of degrees of freedom is time-independent,
an asymptotic ($t \to \infty$) analytic solution to (\ref{kin}) can
be easily found: 
\be
n(x) = \frac{16 \Gamma M_0}{3 m_I^2} x^2
\int_1^\infty\frac{(y-1)^{3/2} dy}{e^{xy}-1}~,
\ee 
where $x=p/T$, and $M_0 \approx M_{\rm Pl}/1.66\sqrt{g^*}$, 
leading to the number density 
\be
N_0=\int \frac{d^3p}{(2\pi)^3}n(p) = 
\frac{3\Gamma M_0 \zeta(5)}{2\pi
m_I^2} T^3
\ee
and to an average momentum of created sterile neutrinos 
$
\langle p \rangle = \pi^6/(378\zeta(5))T = 2.45 T~,
$
which is about $20$\% smaller than that for the equilibrium thermal
distribution, $p_T=3.15 T$.

For the inflaton with a mass $m_I < {\cal O}(500)$ MeV, taking
$g^*=\rm{const}$ is a bad approximation, since exactly in this
region  $g^*$ changes from $g^* \sim 60$ at $T\sim 1$ GeV to $g^*
\sim 10$ at $T\sim 1$ MeV, because of the disappearance of quark and
gluon degrees of freedom. In this case a numerical solution of eq.
(\ref{kin}) is necessary with the input of the hadronic equation of
state, which is not known exactly. To make an estimate we took a
phenomenological  equation of state constructed in \cite{ALS} on the
basis of the hadron gas model at low temperatures, and on available
information on lattice simulations and perturbative computations; we
found that $N= f(m_I) N_0$, where the function  $f(m_I)$ changes from
$0.9$ at $m_I=70$ MeV to $0.4$ at $m_I=500$ MeV. An average momentum
stays almost unchanged in this interval of $m_I$. At higher inflaton
masses a good approximation to $f(m_I)$ is $f(m_I)\simeq
(10.75/g^*(m_I/3))^{3/2}$, and to an average momentum is  $\langle p
\rangle \simeq 2.45 T(10.75/g^*(m_I/3))^{1/3}$.

The abundance of dark matter sterile neutrinos can be further diluted
by a factor $S$, which accounts for the entropy production in decays of
heavier sterile neutrinos  \cite{Asaka:2006ek}.  Collecting all
factors together we get for the contribution of sterile neutrinos to
the dark matter abundance:
\be
\Omega_s \simeq \frac{0.26 f(m_I)}{S}
 \frac{\Gamma M_0 m_s}{m_I^2 \times 12~\rm{eV}}\frac{2\pi
 \zeta(5)}{\zeta(3)}~,
\ee
where $m_s = f_1 \langle \chi \rangle$ is the dark matter sterile
neutrino mass, and $M_0$ is taken with $g^*=10.75$.

Let us proceed to numerical estimates. For the case under
consideration, the Yukawa couplings $F_{\alpha 1}$ are very small and
do not contribute to active neutrino masses \cite{Asaka:2005pn}. So,
the couplings $F_{\alpha 2,3}$ cannot  be smaller than ${\cal
O}(\sqrt{m_{\rm{sol}}M}/v) \simeq 10^{-8}$, where $m_{\rm{sol}}\simeq
0.01$ eV is the solar-neutrino mass difference, $M$ is the mass of
the heavier sterile neutrino, $v$ is the Higgs vev. With the use of
general results of \cite{Asaka:2006ek} one can find that the entropy
production factor can be in the region $1< S < 2$. Requiring that
sterile neutrinos constitute all the dark matter we find that $f_1
\simeq (4$--$5) \times 10^{-11}$ for $m_I \simeq 300$ MeV and that
their mass $m_s$ should be in the interval $m_s \simeq 16$--$20$ keV. 
Quite amazingly, the keV scale for the sterile neutrino mass follows
from observed dark matter abundance and from inflaton self-coupling,
fixed by the observations of fluctuations of the CMB, provided the
inflaton mass is in the GeV region.   The average momentum of sterile
neutrinos, accounting for the dilution factor $S$, can be as small as
$0.6p_T$ in this case.  For the inflaton mass $m_I \sim 100$ GeV, we
find  $f_1 \simeq 10^{-10}$, leading to the sterile neutrino mass
$m_s \sim {\cal O}(10)$ MeV.

A sterile neutrino in this mass range is perfectly consistent with
all cosmological and astrophysical observations. As for the bounds on
mass versus active--sterile mixing coming from X-ray observations of
our galaxy and its dwarf satellites
\cite{Boyarsky:2006fg,Riemer-Sorensen:2006fh}, they are easily
satisfied since the production mechanism of sterile neutrinos
discussed above has nothing to do with the active--sterile neutrino
mixing leading to the radiative mode of sterile neutrino decay. Note
that for small enough $F_{\alpha 1}$ the dark matter sterile neutrino
cannot explain the pulsar kick velocities \cite{Kusenko:1997sp} and
the early reionization\cite{Biermann:2006bu}. As for the limits
coming from Lyman-$\alpha$ forest considerations \cite{Hansen:2001zv}
$m_s > \langle p\rangle/p_T \times m_{\rm{Lyman}}$ where 
$m_{\rm{Lyman}} \simeq 15.4$ keV \cite{Seljak:2006qw}, for $S=2$ one
gets $m_s > 10$ keV. These  values are comfortably within the mass
interval discussed above.

Having fixed the Yukawa coupling constant $f_1$ for the lightest
neutrino, we can estimate the constants $f_2$ and $f_3$. First, these
constants must be nearly equal, so as to achieve the amplification of
CP-violating effects necessary for baryogenesis in the $\nu$MSM
\cite{Asaka:2005pn}. Second, the mass of the heavier sterile
neutrinos should be roughly in the interval $1$--$20$ GeV, the lower
bound comes from the requirement that their decays do not spoil the
big bang nucleosynthesis
\cite{Akhmedov:1998qx,Asaka:2005pn,Asaka:2006ek} while the upper
bound comes from the requirement that the processes with lepton
number non-conservation must be out of thermal equilibrium at the
electroweak scale \cite{Asaka:2005pn}. Depending on the vev of the
inflaton, we arrive at $f_2 \simeq f_3 \lesssim  10^{-4}$. These
values are too small to lead to lepton number violating processes,
which could spoil the baryogenesis mechanism via neutrino
oscillations, provided the decays $\chi \rightarrow 2N_{2,3}$ are
kinematically forbidden. They are also too small to spoil the
flatness of the inflaton potential.

In conclusion, we constructed a minimal model that provides
inflation, gives a candidate for a dark matter particle, explains the
baryon asymmetry of the Universe, being consistent  with neutrino
oscillations. It contains only light particles, in the keV range for
a dark matter sterile neutrino, and in the GeV range for two other
degenerate neutrinos and the inflaton, which makes it to be
potentially testable in laboratory experiments.  The electroweak
scale in the model is related to the vacuum expectation value of the
inflaton. One can go even further and speculate that the Planck scale
may be generated by a similar mechanism. Indeed, if the interaction
of inflaton with gravity is also scale invariant, $L_G =
\frac{1}{g^2} \chi^2 R$, where  $R$ is the scalar curvature, the
Planck scale will be given by  $M_{\rm{Pl}} \sim
\frac{\langle\chi\rangle}{g}$, requiring very small $g \sim
10^{-14}$--$10^{-11}$. If true, the inflation in this theory may
resemble the pre-Big Bang scenario, proposed in
\cite{Veneziano:1991ek}. In this case an estimate of the inflaton
self-coupling $\beta$ may be no longer valid and the values of the
inflaton mass and sterile neutrino mass derived in this paper may be
considered as indicative only. If the requirement of the scale
invariance is given up, the relation between electroweak scale,
sterile neutrino masses and inflaton mass disappears. However,  the
fact  that the sterile neutrino dark matter abundance can be 
determined by the interaction with inflaton rather than with the
fields of the Standard Model remains in force.

%\newpage

%%%%%%%%%%%%%%%%%%%%%%%%%%%%%%%%%%%%%%%%%%%%%%%%%%%%%%%%%%%%%%%%%%%%%%%%
% Acknowledgments
%%%%%%%%%%%%%%%%%%%%%%%%%%%%%%%%%%%%%%%%%%%%%%%%%%%%%%%%%%%%%%%%%%%%%%%%
%{\em Acknowledgments.}---
We thank A. Boyarsky, A. Kusenko and O.Ruchayskiy for discussions.
The work of M.S. was supported in part by the Swiss National Science
Foundation. 

%%%%%%%%%%%%%%%%%%%%%%%%%%%%%%%%%%%%%%%%%%%%%%%%%%%%%%%%%%%%%%%%%%%%%%%%
% References
%%%%%%%%%%%%%%%%%%%%%%%%%%%%%%%%%%%%%%%%%%%%%%%%%%%%%%%%%%%%%%%%%%%%%%%%
%\input{infref.tex}

\end{document}